\documentstyle[preprint,aps]{revtex}   
\begin{document}
\draft

\title{Self-assembled island formation in heteroepitaxial
growth}

\author{Albert-L\'aszl\'o Barab\'asi$^{a)}$}
\address{Department of Physics, University of Notre Dame, Notre
Dame, IN 46556}

\date{\today}

\maketitle

\begin{abstract}
We investigate island formation during heteroepitaxial growth using an
atomistic model that incorporates deposition, activated diffusion and
stress relaxation.  For high misfit the system naturally evolves into
a state characterized by a narrow island size distribution.  The
simulations indicate the existence of a strain assisted kinetic
mechanism responsible for the self-assembling process, involving
enhanced detachment of atoms from the edge of large islands and biased
adatom diffusion.
\end{abstract}



\narrowtext

Heteroepitaxial growth of highly strained structures has gained
interest lately as it offers the possibility to fabricate islands
with
very narrow size distribution without any substrate patterning
process.  Thanks to their small size, these islands, coined
self-assembling quantum dots (SAQD), are  candidates for one
dimensional electron confinement \cite{mrs}.  SAQD formation  has
been observed for a wide range
of
material/substrate combinations, including InAs on GaAs
\cite{mrs,leonard1,miller2,koba,moison,ledentsov}
,
InGaAs on GaAs \cite{leo3,madhukar}, AlInAs on GaAlAs
\cite{Leon},
GeSi on Si \cite{apetz,abstreiter}, InP on InGaP
\cite{seifert}, GaSb on GaAs \cite{hatami} and ZnSe
on
ZnMnSe \cite{xin}, indicating the existence of a not yet
understood
{\it common} mechanism governing the self-assembling process.

In this paper we investigate the kinetics of island formation during
heteroepitaxial growth using a one dimensional model \cite{one-d} that
includes all microscopic elements common to the materials for which
SAQD formation has been observed, namely deposition, activated
diffusion, and strain relaxation at {\it every} deposition and
diffusion event.  Depositing 2ML atoms with lattice constant $a^0_f$
on a substrate with lattice constant $a^0_s$, we find that
sufficiently large misfit, $\epsilon \equiv (a^0_f-a^0_s)/a^0_s$,
leads to self-assembled island formation in the system.  In
particular, $\epsilon=5\%$ and $7.5\%$ leads to a {\it narrowly
peaked} island size distribution, in contrast with a wide distribution
for $\epsilon=0\%$ and $2.5\%$.  We discuss the kinetic mechanism
responsible for the self-assembling process, and compare our results
with experimental work on SAQD formation.  Since the model does not
contain material dependent features, the mechanism responsible for
self-organization is expected to be generic, applicable to a wide
class of materials.

{\it Monte Carlo Method and Stress Relaxation---} 
   To include stress in the
model, we consider that the atoms  interact harmonically
with their nearest
and next nearest neighbors \cite{orr}.   The elastic strain
energies  are 
recalculated  after every deposition or diffusion
event,  using a checkerboard relaxation method. The relaxation
starts at the last active atom, and propagates 
 radially outward,  until in a  two
particle wide shell   all relative displacements are smaller than a
preset  parameter  $\delta$. The relaxation is
iteratively  restarted from the same  origin until even in the
closest shell the displacement is
smaller than $\delta$.  Radial relaxation 
is more  efficient for   our problem than  algorithms  relaxing 
the {\it entire system}, since most  changes in the
displacement
occur  where a  particle moves, and decrease  fast with the
distance from the source.  Trial
runs with different values of  $\delta$ indicated  that for
$\delta
\le 10^{-3}$ the results were practically identical.  In the
simulations we used $\delta=10^{-4}$. 

The surface particles are allowed to hop to neighboring lattice sites,
with an SOS condition, disallowing up or down jumps larger than one
atom high.  The hopping probability for an atom is
proportional to $\exp[-(nE_n + E_0 -E_s)/k_BT]$, where $n$ is the
number of neighboring atoms, $E_n$ is the bond energy and $E_0$ is the
diffusion barrier for an isolated atom on a stress free substrate
\cite{comp}.  The strain energy is given by $E_s=(c/2) \sum (a_i-
a_i^0)^2/(a_i^0)^2$, where $a^0_i$ and $a_i$ are the bulk and
stretched bond lengths, $i$ running over the occupied nearest and
next-nearest neighbors. We use $c=44$eV for the force constant, a
typical value for many semiconductors \cite{harris}, $E_n=0.3$eV,
$E_0=0.4$eV and T=800K.  The substrate consists of $N=50$ML atoms with
lattice constant $a_s^0=1$, on which we deposit with a {\it constant
flux} 2ML atoms with lattice constant $a_f^0=a_s^0(1+\epsilon)$. The
system size is $L=200$ \cite{comm1}.  We identify as island every
mound with height larger than 1ML, and define the {\it base size of
the island}, denoted by $s$, as the lateral size of the island
measured in the {\it second monolayer} (to distinguish it from the
wetting layer). The islands are coherently strained and dislocations
are {\it not} allowed.

{\it Numerical results---} The most convincing evidence of the
stress
induced self-assembling process is provided by the island size
distribution, shown in Fig. \ref{fig-distr}a. For $\epsilon=0\%$
and
$2.5 \%$ the distribution is wide, i.e. the system contains
islands of
all sizes, with a small peak around $s=20$. However, for
$\epsilon=5\%$ and $ 7.5 \%$ the distribution has a {\it narrow
peak }
centered at $s=6$ for $\epsilon=5\%$ and $s=5$ for
$\epsilon=7.5\%$.

The parameter capturing the dynamics of self-assembly in the system is
the {\it relative width}, $w_s/{\bar s}$, shown in Fig.
\ref{fig-distr}a, where $w_s^2 \equiv \bar{s^2}-{\bar s}^2$ is the
width of the island size distribution and $\bar s$ is the average
island size. An increasing $w_s/{\bar s}$ indicates unbounded growth
of fluctuations, while a {\it decreasing} one is a signal of
self-organization in the system.  As Fig. \ref{fig-distr}a indicates,
for $\epsilon=0\%$ and $2.5 \%$ $w_s/{\bar s}$ increases continuously
with coverage, while for $\epsilon=5\%$ and $7.5\%$ $w_s/{\bar s}$
increases only until it reaches a peak at some small coverage
$\theta_c$, after which it decays.  The peak signals the onset of
self-organization: for $\theta > \theta_c$ we witness a continuous
increase in the {\it uniformity} of the island size.  The peak is at
$\theta_c=0.66$ML for $\epsilon=5\%$, and $\theta_c=0.15$ML for
$\epsilon=7.5\%$, indicating that the self-assembling process is more
effective for larger misfit.

An experimentally often measured quantity is the island density,
$\rho$.  As Fig. \ref{fig-distr}b indicates, for the stress free
system $\rho$ has a peak at $1.33$ML, after which it decreases.
This
behavior is  characteristic for  homoepitaxy \cite{barabasi95}:
for
small
coverages the dynamics is dominated by island nucleation. After a
certain $\rho$ is reached the incoming atoms are captured by the
existing islands, prohibiting further island nucleation, and
stabilizing the island density.  Continuing the deposition leads
to
island coalescence, that results in a drop of the island density.
Fig. \ref{fig-distr}b is consistent with this scenario for
$\epsilon=0\%$ and $\epsilon=2.5 \%$. However, we observe no such
peak
for $\epsilon=5\%$ and $7.5\%$, indicating {\it continuous island
nucleation}, without significant coalescence \cite{comm3}.

{\it Mechanism of self-organization---} The main difference
between
the
stress free  and the stressed system  comes in  two strain
related effects, that  we discuss separately. 

(a) Strain lowers the energy barrier for diffusion, thus making
diffusive hops more probable.  Fig. \ref{fig-strain} shows the strain
energy in the vicinity of an island for $\epsilon = 7.5\%$, indicating
that the {\it substrate is strained} and that $E_s$ decreases as we
move away from the edge of the island. This means that if atoms are
deposited near the island, strain biases their otherwise random
motion, generating a net surface current ${\bf j} = -{\bf \nabla}
\mu(x)$, where $\mu(x)$ is the local chemical potential
\cite{barabasi95}. The chemical potential is $\mu \simeq
-(nE_n+E_0-E_s)$, where $nE_n+E_s$ is independent of the atom position
for an isolated adatom on a flat surface. The only contribution to the
current comes from the position dependence of the strain energy,
leading to ${\bf j} \simeq - \nabla E_s$, that points towards the
decreasing strain direction.  Thus the strain field around an island
generates a {\it net current of adatoms away from the island}.

(b) For large  islands the strain energy, $E_s$, at the edge
becomes comparable to the bonding energy of the edge atom, $nE_n
+
E_0$ (with $n=1$), enhancing its detachment, thus leading to a 
gradual dissolution of the island. Such mechanism favors a
smaller
average island size and leads to a narrower island size
distribution,
as observed by Ratsch {\it et al.} \cite{Ratsch94}.

The simultaneous action of (a) and (b) leads to a kinetic
mechanism
stabilizing the island size: as islands grow, a strain field
develops, that helps to dissolute the edge atoms (effect (b))
and
"pushes" them away from the islands (effect (a)).  Furthermore,
the
newly deposited atoms also diffuse away from the larger islands
(effect (a)).  These combined effects  {\it slow the growth rate
of
large islands} and {\it increase the adatom density} away from
them,
thus {\it enhancing the nucleation of new islands}. The newly
nucleated islands are small, and so is the strain field around
them,
thus they grow at a much faster rate than the older and larger
one.
This  eventually leads to a narrow island size distribution in
the system (Fig. \ref{fig-distr}a).  The final island size is
determined by a {\it
dynamical  balance} between the adatom density, the strain
induced
current away from the islands and the strain energy of the
island,
governing the detachment of the edge atoms.  If we could monitor
in
real time the growth,  we would witness a
continuous
nucleation of islands, such that small islands grow fast, and
stop
growing after they reach a certain size.

{\it Discussion---} The observed behavior compares favorably with the
main features of the experimentally observed SAQD formation.  First,
TEM observations of the strained islands document the existence of the
strain field in the substrate \cite{leo3,madhukar,ponchet}. As
Fig. \ref{fig-strain} shows, such a field is reproduced by our
relaxation method, and is responsible for the current ${\bf j}$
discussed in (a).  Second, it is experimentally established that the
increasing coverage contributes mainly to an increase in the island
density, and much less to the further increase in the size of the
existing islands \cite{leonard1}, which is reproduced by the
simulations (see Fig. \ref{fig-distr}b).  Third, experiments on GaInAs
growth on GaAs substrate document an increasing width in the early
stages of the deposition process, followed by a gradual decay for
larger coverages \cite{leonard1,koba}. This is similar to the behavior
shown in Fig. \ref{fig-distr}a: $w/{\bar s}$ decreases only after a
certain coverage $\theta_c$ has been reached. Indeed, for small
coverages the islands are both small and distant, thus the strain
induced biased diffusion (a) and atom detachment (b) are not yet
relevant, and the island formation essentially follows a strain-free
path.  Only when $\theta$ approaches $\theta_c$ the discussed
strain-induced mechanisms reverse the growth in the relative
width. Finally, the simulations indicate ordering in the distances
between the islands, as observed for high island densities in some
investigations \cite{ledentsov}.

 I have benefited from enlightening  discussions  with J.K.
Furdyna, M. Krishnamurthy, N.N. Ledentsov, J.L. Merz, M.S. 
Miller,
K. Newman, and  W. Seifert.

\begin{figure}
\caption{ (a) Island size distribution measured after the
deposition
of 2ML atoms.  Inset:  Relative width $w_s/{\bar s}$ as a function of
coverage. (b) Island density as a function of coverage.  In all
figures the symbols correspond to misfit values 0$\%$ ($\circ$),
2.5$\%$ ($\Box$), 5$\%$ ($\Diamond$), and 7.5$\%$ ($\triangle$).}
\label{fig-distr}
\end{figure}

\begin{figure}
\caption{The strain energy around a typical island. The substrate
(filled $\Box$) and the islands on top of it (empty $\Box$) are
shown
on the upper part of the figure. $E_s$ is the strain energy of an
atom
placed on the top of the substrate or on the island. For example,
$E_s$ at $x=18$ is the strain energy felt by the adatom
shown by the
circle on the top of the substrate.  One can see that $E_s$
is
the largest when the atom is at the edge of the island
($x=21,~30$).
$E_s$ decays as the adatom moves away from the island, generating
a
net current, ${\bf j}(x)$, shown by the arrows. Note that $E_s$
does
not decay to zero, since the monomer can locally stretch the
substrate.}
\label{fig-strain}
\end{figure}



\begin{references}

\bibitem[a)]{byline} Email address: alb@nd.edu



\bibitem{mrs} For a recent review see W. Seifert, N. Carlsson,
M-E. Pistol, L. Samuelson, and L.R. Wallenberg, J.  Progr. Crystal
Growth Charact. Mater. {\bf 33} 423 (1996);  P.M. Petroff, and
G. Medeiros-Ribeiro, MRS Bulletin {\bf 21} (No 4), 50 (1996).




\bibitem{leonard1}
D. Leonard, K. Pond, and P.M. Petroff, Phys. Rev. B {\bf 50},
11687 (1994).



\bibitem{miller2}
M.S. Miller,
S. Jeppesen, D. Hessman, B. Kowalski, I. Maximov, 
B. Junno, and L. Samuelson, 
Solid-State Electr. {\bf 40}, 609 (1996).

\bibitem{koba}
N.P. Kobayashi, T.R. Ramachandran, P.Chen, and A. Madhukar, Appl.
Phys. Lett. {\bf 68}, 3299 (1996).

\bibitem{moison}
J.M. Moison, F. Houzay, F. Barthe, L. Leprince, E. Andr\'e, and
O. Vatel,
 Appl. Phys. Lett. {\bf 64}, 196 (1994).

\bibitem{ledentsov} S. Ruvimov, P. Werner, F. Hatami,
K. Scheerschmidt, J. Heydenreich, U. Richter, N.N. Ledentsov,
M. Grundmann, D. Bimberg, V.M. Ustinov, A. Yu. Egorov, P.S. Kop'ev,
and Zh. I. Alferov, Phys. Rev. B {\bf 51}, 14766 (1995).


\bibitem{leo3} D. Leonard,
M.Krishnamurthy, C.M. Reaves, S.P. Denbaars, and P.M. Petroff,
Appl. Phys. Lett. {\bf 63}, 3203 (1993).

\bibitem{madhukar}
A. Madhukar,
Q. Xie, P. Chen, and A. Konkar,
Appl. Phys. Lett. {\bf 64}, 2727 (1994).

\bibitem{Leon} R. Leon, S. Fafard, D. Leonard, J.L. Merz, and P.M. Petroff, 
Appl. Phys. Lett. {\bf 67}, 521 (1995).

\bibitem{apetz} R. Apetz, L. Vescan, A. Hartman, C. Diecker, and
H. Luth, Appl. Phys. Lett. {\bf 66}, 445 (1995); P. Schittenhelm,
M. Gail, J. Brunner, J.H. Nutzel, and G. Abstreiter, {\it ibid.} {\bf
66} 1292 (1995); M.Krishnamurthy, J.S. Drucker, and J.A. Venables,
J. Appl. Phys. {\bf 69}, 6461 (1991); D.E. Jesson, K.M. Chen, and
S.J. Pennycook, MRS Bulletin {\bf 21} (No 4), 31 (1996).

\bibitem{abstreiter}
G. Abstreiter, 
P. Schittenhelm, C. Engel, E. Silveira, 
A. Zrenner, D. Meertens, and W. Jager, 
Semic. Sci. Techn. (in press).

\bibitem{seifert} N. Carlsson, W. Seifert, E. Peterson, M. Castillo,
 M.E. Pistol, and L. Samuelson, Appl. Phys. Lett. {\bf 65}, 3093
 (1994).



\bibitem{hatami} F. Hatami, N.N. Ledentsov, M. Grundmann, J. Bohrer,
F. Heinrichsdorff, M. Beer, D. Bimberg, S.S. Ruminov, P.Werner,
U. Gosele, J. Heydenreich, U. Richter, S.V. Ivanov, B. Ya. Meltser,
P.S. Kop'ev, and Zh. I. Alferov, Appl. Phys. Lett. {\bf 67}, 656
(1995).

\bibitem{xin} S.H. Xin, P.D. Wang, A. Yin, M. Dobrowolska, J.L.  Merz,
and J.K. Furdyna, Appl. Phys. Lett. {\bf 69}, 3884 (1997).
 

\bibitem{one-d} The fact that the model is one dimensional is not
expected to affect the nature of the self-organization process, but it
can modify such quantities as the island size and critical coverage.


\bibitem{orr}
B.G. Orr, D. Kessler, C.W. Snyder, and L.M. Sander, 
Europhys. Lett. {\bf 19}, 33 (1992).




\bibitem{comp} Note that compressive (tensile) strain lowers
(incerases) the energy barrier for diffusion [see M. Schroeder and
D.E. Wolf (preprint)]. For SAQD formation the strain is compressive,
thus we use a negative sign for the strain energy. This sign also
determines the direction of the surface current near an island
discussed under (a).

\bibitem{harris} W.A. Harrison, {\it Electronic Structure and the
Properties of Solids} (Freeman, San Francisco, 1980).

\bibitem{comm1} The time consuming strain relaxation limits the 
diffusion length, $\ell_d$.  The typical $\ell_d$ in our
simulations varies
between 19 lattice sites for $\epsilon = 0\%$ to 31 for $\epsilon
=
7.5\%$. This is much shorter than the experimental diffusion
length,
which is of order of hundreds of nm.  This decreases the average
distance between the islands and their average size, since there
will
be a smaller surface area from where the islands can collect
atoms. Physically, the curtained $\ell_d$ corresponds to growth
with
high flux.

\bibitem{comm3} Note that in Fig. \protect\ref{fig-distr}b $\rho$
increases much faster for $\epsilon=7.5\%$ than for
$\epsilon=5\%$. While for $\epsilon=5\%$ a wetting layer is formed,
for $\epsilon=7.5\%$ all deposited atoms aggregate in islands. Thus we
suspect that for $\epsilon=7.5\%$ the strain energy is large enough to
lead to the Volmer-Weber growth mode (3D island formation), in
contrast with $\epsilon \le 5\%$, that grows in the Stranski-Krastanov
growth mode. The detailed account of this transition will be presented
elsewhere.


\bibitem{barabasi95} For a recent review see A.-L. Barab\'asi and
H. E. Stanley, {\it
Fractal
Concepts in Surface Growth} (Cambridge University Press,
Cambridge,
1995).


\bibitem{Ratsch94} C. Ratsch, A. Zangwill, and P. Smilauer, Surf.
Sci. {\bf 314},  L937 (1994).




\bibitem{ponchet} A. Ponchet, A. Le Corre, H. L'Haridon, B. Lambert,
and S. Sala\"un, Appl. Phys. Lett. {\bf 67}, 1850 (1995).







\end{references}
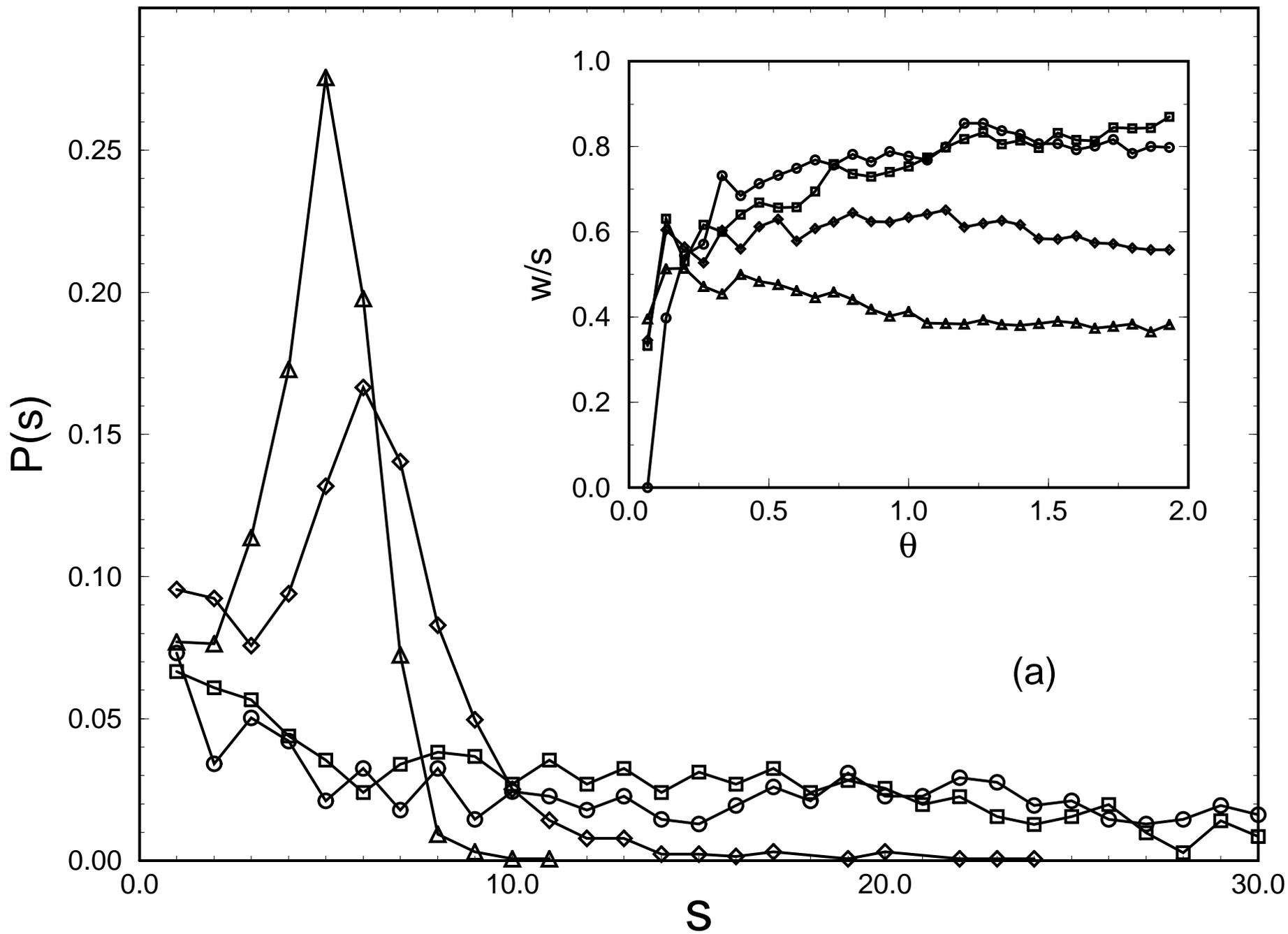
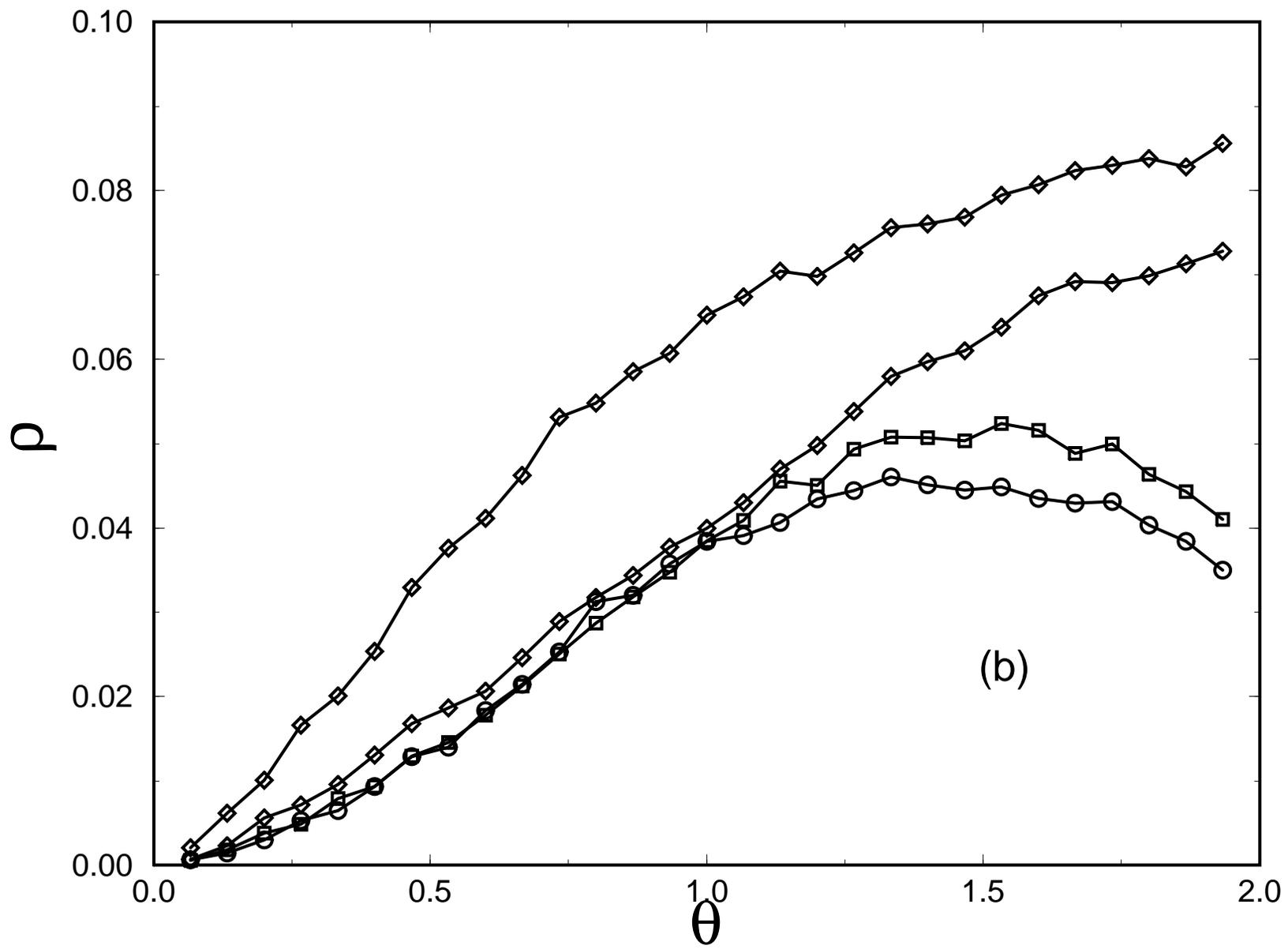
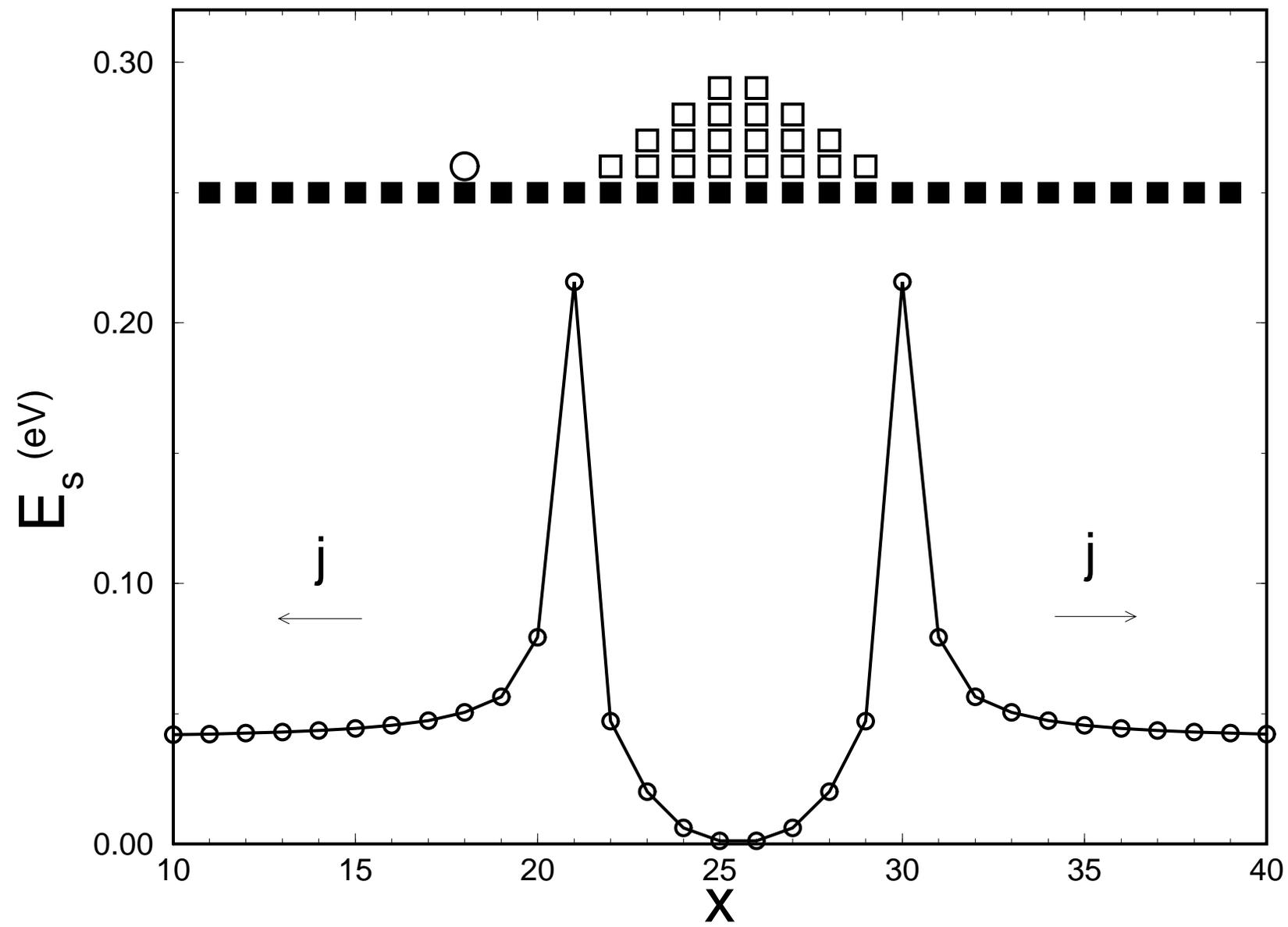
\end{document}